\documentclass[11pt,a4paper]{elsarticle}

\usepackage[T1]{fontenc}
\usepackage{hyperref}
\hypersetup{
    colorlinks=true,
    linkcolor=blue,
    filecolor=blue,
    urlcolor=blue,
    }

\usepackage{geometry}
\usepackage{graphicx}
\usepackage{tikz}
\usepackage[labelfont=bf]{caption}
\usepackage{subcaption}
\usepackage{booktabs}
\usepackage{amsfonts}
\usepackage{dsfont}
\usepackage{bm}
\usepackage{comment}
\usepackage{amssymb}
\usepackage{amsmath}

\makeatletter
\def\ps@pprintTitle{%
  \let\@oddhead\@empty
  \let\@evenhead\@empty
  \def\@oddfoot{\reset@font\hfil\thepage\hfil}
  \let\@evenfoot\@oddfoot
}
\makeatother

\renewcommand{\figurename}{{Fig.}}

\begin{document}

\begin{frontmatter}

\title{Structural prediction of super-diffusion in multiplex networks}

\author[inst1]{Llu\'is Torres-Hugas}
\author[inst1]{Jordi Duch}
\author[inst1]{Sergio G\'omez}

\affiliation[inst1]{
            organization={Departament d'Enginyeria Inform\`atica i Matem\`atiques, Universitat Rovira i Virgili},
            city={Tarragona},
            postcode={43007},
            country={Spain}}

\begin{abstract}
Diffusion dynamics in multiplex networks can model a diverse number of real-world processes. In some specific configurations of these systems, the super-diffusion phenomenon arises, in which the diffusion is faster in the multiplex network than in any of its layers. Many studies attempt to characterize this phenomenon by examining its dependency on structural properties of the network, such as overlap, average degree, network dissimilarity, and others. While certain properties show a correlation with super-diffusion in specific networks, a broader characterization is still missing. Here, we introduce a structural parameter based on the minimum node strength that effectively predicts the occurrence of super-diffusion in multiplex networks. Additionally, we propose a novel framework for deriving analytical bounds for several multiplex networks structures. Finally, we analyze and justify why certain arrangements of the inter-layer connections induce super-diffusion. These findings provide novel insights into the super-diffusion phenomenon and the interplay between network structure and dynamics.
\end{abstract}


\end{frontmatter}


\section{Introduction}
\label{sec:introduction}

Diffusion dynamics in complex networks has garnered significant research interest due to its ability to model numerous behaviors in social and transportation networks \cite{brockmann2008anomalous,de2014navigability}, epidemiology \cite{saumell2012epidemic}, and biological systems \cite{west1997general}, among others \cite{moreno2004dynamics,buldyrev2010catastrophic,masuda2017random}. In this context, a diffusion process involves the propagation of an entity across connected nodes within a network and is characterized by the diffusion time; the smaller its value, the faster its convergence to the equilibrium state. The structure of the network plays a major role in the dynamics of these systems, so one of the key challenges for predicting their possible outcomes is understanding this complex interplay.

In practice, many real-world networks tend to be more complex than single-layer network structures. For example, the transportation network of a city can consist of interactions between different transportation systems, such as bus and subway networks. In these scenarios, the system is better described by multilayer networks \cite{kivela2014multilayer}, whose research interest arose mainly due to its new emerging properties \cite{gao2012networks,gomez2013diffusion,sole2016congestion}. A particular type of multilayer network with great research interest is the multiplex network \cite{de2013mathematical}, in which each node is present in all the layers and the inter-layer links only connect instances of the same node in different layers.

In~\cite{gomez2013diffusion}, an interesting behavior was shown for diffusive processes in multiplex networks: in some specific cases, the multiplex network diffusion time can be smaller than in any of its (isolated) layers. This phenomenon is known as \textit{super-diffusion}. Since then, many attempts have been made to understand and characterize it. Early works suggested that network dissimilarity can enhance the occurrence of super-diffusion \cite{serrano2017optimizing}, while others have found that similar intra-layer diffusion coefficients and low overlap also play an important role \cite{cencetti2019diffusive}. Recent articles have also focused on the importance of the average degree of the layers and how certain arrangements of the inter-layer connections between nodes can impact the phenomenon \cite{cai2023occurrence}. In particular, they observed that in some duplex networks (i.e., two-layer multiplex networks), Negatively Correlated inter-layer connections (NC) are more beneficial to the occurrence of super-diffusion than Positively Correlated (PC) ones. A NC (PC) duplex network is built by establishing the inter-layer links in a maximally disassortative (assortative) way, namely: sorting the nodes of the first layer by increasing strength, sorting the nodes of the second layer by decreasing (increasing) strength, and then connecting the first nodes in both layers, next the second nodes, and so on until each node is connected to its equally ranked counterpart. Despite the efforts, a more general characterization is still lacking.

In this work, we propose a general structural parameter based on the minimum node strength to predict the occurrence of super-diffusion in multiplex networks. This parameter can also justify the previous observations that a NC configuration potentiates the appearance of super-diffusion. Furthermore, we introduce a novel approach to derive analytical bounds for some multiplex networks structures, with potential generalization to other network configurations. Finally, we hypothesize and analyze that, in general, a NC configuration of the inter-layer connections is the best way to induce super-diffusion in multiplex networks.

The outline of this work is as follows. In Section~\ref{sec:model} we present the theoretical framework of diffusion in multiplex networks. We then introduce the structural parameter and test its accuracy in Section~\ref{sec:parameter}. Next, in Section~\ref{sec:bounds}, we derive theoretical bounds for super-diffusion for three specific configurations. In Section~\ref{sec:maximum} we justify our hypothesis that a NC configuration is the best super-diffusion enhancement strategy, and finally, in Section~\ref{sec:conclusions}, we summarize the main contributions and discuss some extra remarks and implications.

\section{Diffusion and super-diffusion in multiplex networks}
\label{sec:model}

\begin{figure}[tb!]
  \begin{center}
  \begin{tabular}[t]{lll}
     RA & PC & NC
     \\
     \includegraphics[width=0.32\textwidth]{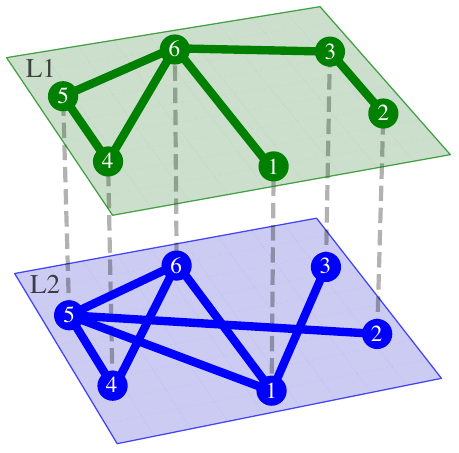}
     &
     \includegraphics[width=0.32\textwidth]{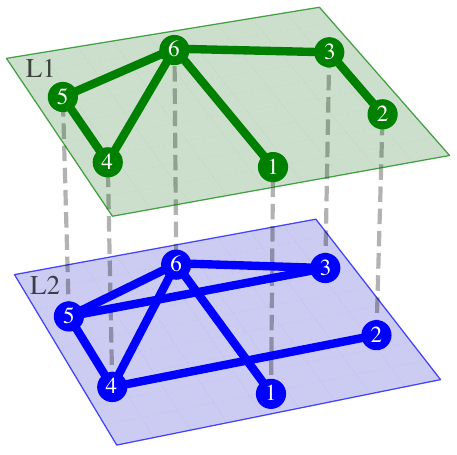}
     &
     \includegraphics[width=0.32\textwidth]{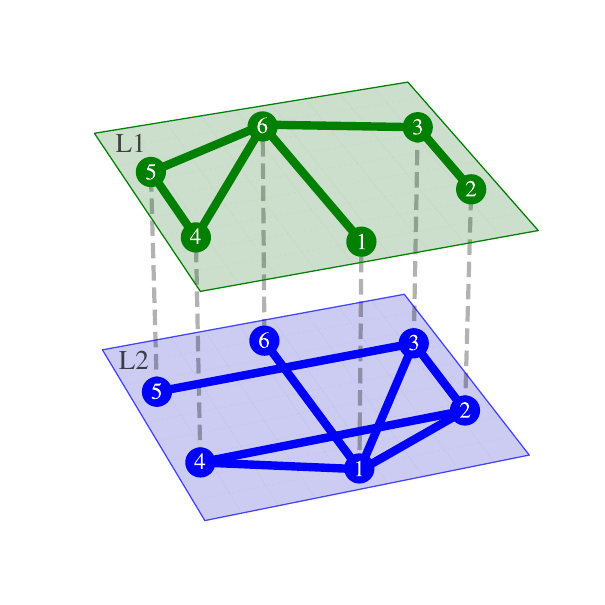}
    \\ \\
    \multicolumn{3}{c}{
    \begin{tabular}[t]{cccccccc}
      \toprule
      Node Label & & 1 & 2 & 3 & 4 & 5 & 6
      \\ \hline \hline
      L1 degrees &   & 1 & 1 & 2 & 2 & 2 & 4
      \\ \hline
      & RA & 3 & 1 & 1 & 2 & 4 & 3
      \\ \cline{2-8}
      L2 degrees & PC & 1 & 1 & 2 & 3 & 3 & 4
      \\ \cline{2-8}
      & NC & 4 & 3 & 3 & 2 & 1  & 1
      \\
      \bottomrule
    \end{tabular}
    }
  \end{tabular}
  \end{center}
  \caption{Three multiplex networks formed by the same two layers, each with 6~nodes, and with Randomly Assigned (RA), Positively Correlated (PC), and Negatively Correlated (NC) inter-layer connections, respectively. Additionally, we display a table with the corresponding node degree associations for the considered inter-layer connections. Note that the degrees are increasing for L1 and L2-PC, and decreasing for L2-NC.}
  \label{fig:diagram}
\end{figure}

Following \cite{gomez2013diffusion}, assume we have a multiplex network composed with $M$~layers and $N$~nodes per layer. In each layer ${\alpha = 1,\dots,M}$, we assign a label ${i=1,\dots,N}$ to each node, and nodes with the same label across different layers are interconnected, see Fig.~\ref{fig:diagram}.

The diffusion dynamics in each layer~$\alpha$ is governed by a diffusion constant~$D_\alpha$, and the diffusion between different layers~$\alpha$ and~$\beta$ by a parameter~$D_{\alpha \beta}$. The dynamical evolution equation of state $x^{[\alpha]}_i$ of node~$i$ in layer~$\alpha$ is
\begin{equation} \label{eqdiff}
    \frac{dx_i^{[\alpha]}}{dt} = D_\alpha \sum_{j=1}^N w_{ij}^{[\alpha]} (x_j^{[\alpha]} - x_i^{[\alpha]}) + \sum_{\beta = 1}^M D_{\alpha \beta} (x_i^{[\beta]} - x_i^{[\alpha]})\,,
\end{equation}
where $w_{ij}^{[\alpha]}\geqslant 0$ denotes the elements of the weight matrix at layer~$\alpha$, which are zero if there is no link between nodes~$i$ and~$j$ in layer~$\alpha$. The set of Eqs.~\eqref{eqdiff} can be written in matrix form as
\begin{equation} \label{eqsist}
    \bm{\Dot{x}} = -\mathcal{L}^{\mathcal{M}} \bm{x}\,,
\end{equation}
where $\bm{x} \in \mathds{R}^{N M}$ is the state vector of all nodes in all layers, and $\mathcal{L}^{\mathcal{M}}$ is the supra-Laplacian matrix of the multiplex network, which for the duplex case $M=2$ takes the form
\begin{equation}
    \mathcal{L}^{\mathcal{M}} =
    \begin{pmatrix}
    D_1 L_1 + D_{12} I & - D_{12} I \\
    - D_{21} I &  D_2 L_2 + D_{21} I
    \end{pmatrix},
\end{equation}
where $L_1$ and $L_2$ are the Laplacian matrices of each layer, and $I$ is the $N\times N$ identity matrix. The Laplacian matrix of each layer $\alpha$ is $L_\alpha = S_\alpha - W_\alpha$, where $W_\alpha$ is the weights matrix and $S_\alpha$ is the diagonal matrix containing the strength of each node $i$ at layer $\alpha$, i.e., $s_{i}^{[\alpha]} \equiv s_{ii}^{[\alpha]} = \sum_{j} w_{ij}^{[\alpha]}$.

The general solution to Eq.~\eqref{eqsist} is
\begin{equation} \label{eqdiffsol}
  {\bm{x}(t) = \sum_{k} C_k e^{-\lambda^{\mathcal{M}}_k\, t} \bm{v}_k}\,,
\end{equation}
where $\{\lambda^{\mathcal{M}}_k\}$ and $\{\bm{v}_k\}$ are the eigenvalues and eigenvectors of the supra-Laplacian matrix, respectively, and $\{C_k\}$ are constants that could be determined from the initial condition. Supposing that the layers are connected undirected networks, that $D_{\alpha \beta} = D_{\beta \alpha}$ for all pairs of different layers, and making use of the property that in this case the Laplacians are positive semi-definite, we may sort the eigenvalues as $0 = \lambda_1^{\mathcal{M}} < \lambda_2^{\mathcal{M}} \leqslant \lambda_3^{\mathcal{M}} \leqslant \cdots \leqslant \lambda_{N M}^{\mathcal{M}}$. The convergence of Eq.~\eqref{eqdiffsol} to the equilibrium is governed by the smallest non-zero eigenvalue $\lambda_2^{\mathcal{M}}$, which determines the diffusion time $\tau^{\mathcal{M}} = 1 / \lambda_2^{\mathcal{M}}$. Therefore, the larger $\lambda_2^{\mathcal{M}}$, the faster the convergence to the equilibrium state. From now on, without loss of generality, we assume $D_{\alpha \beta} = D_x$ $\forall \alpha, \beta$ ($\alpha \neq \beta)$ and $D_\alpha = 1$ $\forall \alpha$.

A multiplex network is said to be super-diffusive if the diffusion time in the multiplex network is smaller than the diffusion time in all layers:
\begin{equation}
    \tau^{\mathcal{M}} < \min_\alpha (\tau^{[\alpha]})\,,
\end{equation}
or equivalently,
\begin{equation}
    \lambda_2^{\mathcal{M}} > \max_\alpha (\lambda_2^{[\alpha]})\,,
\end{equation}
In the limit $D_{x} \rightarrow \infty$, the value of $\lambda_2^{\mathcal{M}}$ is equal to the second smallest eigenvalue of the average Laplacian $L^{\mathcal{A}} = \frac{1}{M} \sum_{\alpha} L_\alpha$, i.e., $\lambda_2^{\mathcal{M}}=\lambda_2^{\mathcal{A}}$, see \cite{sole2013spectral}. Note that $L^{\mathcal{A}}$ is the Laplacian of the average superposition network (ASN), $W^{\mathcal{A}} = \frac{1}{M} \sum_{\alpha} W_\alpha$; for duplex networks, it corresponds to the semi-sum of the layers, $W^{\mathcal{A}} = \frac{1}{2}(W_1+W_2)$ \cite{gomez2013diffusion}. Since $\lambda_2^{\mathcal{M}}$ monotonically increases with $D_x$, a multiplex network cannot be super-diffusive unless it is super-diffusive at $D_{x} \rightarrow \infty$. Therefore, we just need to compare the second smallest eigenvalue of the Laplacian of the ASN with that of the individual layers to know if a multiplex network is super-diffusive (for large enough values of $D_x$). In the rest of this paper, we assume we are in the limit $D_{x} \rightarrow \infty$.

A simple indicator of the super-diffusion intensity was proposed in \cite{cencetti2019diffusive}:
\begin{equation} \label{sdintensity}
    \xi = \frac{\lambda_2^{\mathcal{A}} - \max_\alpha (\lambda_2^{[\alpha]})}{\max_\alpha (\lambda_2^{[\alpha]})}\,.
\end{equation}
When $\xi > 0$, the multiplex network is super-diffusive. An important remark of the super-diffusion intensity is that, if $\xi_1 > \xi_2$ for two structurally distinct systems, it does not imply that diffusion is faster in the system with a higher intensity value, or that the occurrence of super-diffusion is more likely in randomly generated systems under the same structural parameters, but rather that the multiplex structure is more beneficial to the diffusion dynamics.

\section{Characterizing super-diffusion with minimum node strength}
\label{sec:parameter}

The second smallest eigenvalue of networks, also named the algebraic connectivity or Fiddler value, has been largely studied both in graph theory \cite{van2023graph,fiedler1973algebraic,jamakovic2008robustness,mckay1981expected} and network science \cite{sole2013spectral,restrepo2006characterizing,jiang2023searching,li2018maximizing}, but it is still unclear how, in general, the structure of the network determines its value. There exists a well-known upper bound \cite{fiedler1973algebraic}:
\begin{equation} \label{fidler}
    \lambda_2 \leqslant \frac{N}{N-1} s_\text{min}\,,
\end{equation}
where $s_\text{min}$ is the minimum node strength value in the network. Other studies \cite{li2018maximizing,jamakovic2008robustness} suggest that $s_\text{min}$, or equivalently the minimum node degree $k_\text{min}$ in unweighted networks, plays a pivotal role in determining the value of $\lambda_2$. In addition, one can easily demonstrate that the inequality in Eq.~\eqref{fidler} becomes an equality for Fully Connected (FC) networks in which all nodes possess equal strength~$s$:
\begin{equation} \label{fc}
    \lambda^{\mathrm{FC}}_2 = \frac{N}{N-1} s\,.
\end{equation}

Our goal is to understand what are the structural conditions that lead to super-diffusion in multiplex networks. This apparent correlation with the minimum node strength has potential to aid our efforts. Furthermore, it can offer insights into why, as observed in many multiplex network systems, certain correlations between degrees of nodes in the inter-layer connections induce certain dynamical properties \cite{kumar2017mirror,cai2023occurrence}.

If we assume that there exists some unknown but almost direct relationship $\lambda_2 \sim f(s_\text{min})$, we can define the following structural parameter for multiplex super-diffusion, inspired by Eq.~\eqref{sdintensity}:
\begin{equation}
    \eta = \frac{s_\text{min}^{\mathcal{A}} - \max_\alpha (s_\text{min}^{[\alpha]})}{\max_\alpha (s_\text{min}^{[\alpha]})}\,,
\end{equation}
where $s_\text{min}^{\mathcal{A}}$ and $s_\text{min}^{[\alpha]}$ are the minimum node strengths of the ASN and of the layer $\alpha$, respectively. When $\eta > 0$, the minimum node strength of the ASN is larger than that of any of the (isolated) layers, and in this case, we predict super-diffusion; otherwise, if $\eta < 0$, we predict the absence of super-diffusion. In the case of $\eta = 0$, we will also predict super-diffusion because, as we will see in Section~\ref{sec:maximum}, the ASN distributes the strength across more links and, in general, this is beneficial to increase~$\lambda_2^{\mathcal{A}}$. This matches very well with the observation that a NC configuration enhances super-diffusion, because what it is actually doing is maximizing the value of $s_\text{min}^{\mathcal{A}}$, and thus, if our assumption is correct, increasing $\lambda_2^{\mathcal{A}}$.

To analyze and test the capacity of parameter~$\eta$ in predicting super-diffusion, we first consider duplex networks formed by two unweighted Erd\H{o}s-R\'enyi~(ER) layers, for a large range of values of the average degrees~$\langle k^{[\alpha]} \rangle$ of the two layers, and for three different arrangements of the inter-layer connections: Randomly Assigned (RA), Positively Correlated (PC), and Negatively Correlated (NC). We generate 50~duplex networks for each pair of the connection probabilities~$p^{[\alpha]}$ of the ER layers, and calculate the accuracy of the prediction~($\Delta$) as the ratio of correctly predicted presence or absence of super-diffusion, and the super-diffusion probability~($q$) as the fraction of super-diffusive duplex networks, for the given structural parameters of the layers.

\begin{figure}[tb!]
\includegraphics[width=1.0\textwidth]{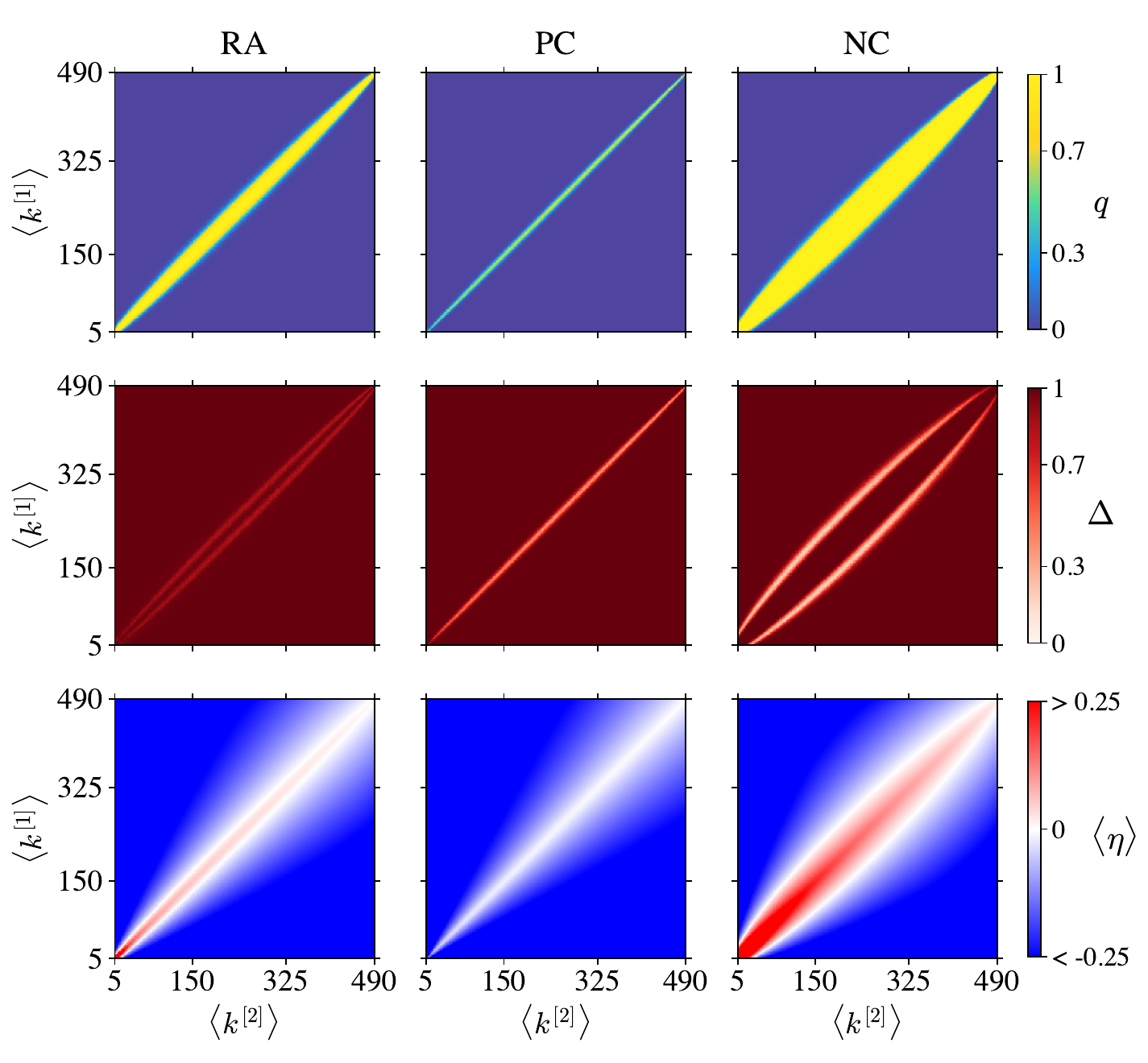}
  \caption{Super-diffusion probability~$q$ (top row), prediction accuracy~$\Delta$ (middle row), and average structural parameter~$\langle \eta \rangle$ (bottom row) as a function of the average degrees of the layers, for duplex networks composed by two unweighted ER layers and with $N=500$. Each column corresponds to one of the three different models of inter-layer connection: RA, PC, and NC. We compute 50~different duplex networks for each combination of $p^{[1]}$ and $p^{[2]}$, with these connection probabilities ranging from~$5/(N-1)$ to~$1$ in steps of~$1/(N-1)$, and ensuring that the layers are connected networks.}
\label{fig:test}
\end{figure}

In Fig.~\ref{fig:test} we show, in three separate rows, the super-diffusion probability~$q$, the prediction accuracy~$\Delta$, and the average structural parameter~$\langle \eta \rangle$ for each pair of average degrees $\langle k^{[1]} \rangle$ and $\langle k^{[2]} \rangle$ of ER-ER duplex networks. The columns correspond to the three considered types of inter-layer connections. We notice that PC inter-layer connections decrease the occurrence of super-diffusion as compared to RA, whereas NC inter-layer connections lead to a higher occurrence of super-diffusion.

If we now consider other duplex structures, for example, unweighted scale-free (SF) layers with exponent $\gamma = 3$ built using the configuration model, we can check whether this approach generalizes to other structures. In order to quantify the general performance, we calculate the global accuracy ($\widehat{\Delta}$) for each duplex layout, which indicates the average accuracy over all considered structural parameters.

\begin{table}[tb!]
\centering
\begin{tabular}[t]{ccccc}
\toprule
L1 & L2 & Inter-layer Connections & Super-diffusion Ratio (\%) & $\widehat{\Delta}$ (\%) \\
\midrule
   &    & RA & 26.7 & 98.0 \\
ER & ER & PC & \phantom{0}5.1 & 96.4 \\
   &    & NC & 49.5 & 87.0 \\ \hline
   &    & RA & 36.3 & 99.0 \\
SF & SF & PC & \phantom{0}3.4 & 99.0 \\
   &    & NC & 85.0 & 93.1 \\ \hline
   &    & RA & 26.2 & 97.7 \\
ER & SF & PC & \phantom{0}3.6 & 97.9 \\
   &    & NC & 59.8 & 88.1 \\
\bottomrule
\end{tabular}
\caption{Ratio of explored space with super-diffusion, and global accuracy of the super-diffusion predictions, for duplex networks with $N=500$, $\langle k^{[1]} \rangle, \langle k^{[2]}\rangle \in [5,100]$, for the three possible combinations of ER and SF layers, and with the three different types of inter-layer connections.}
\label{table:test}
\end{table}

In Table~\ref{table:test} we can see that, overall, the global accuracy of the prediction is really good, and even though they are not perfect, this method gives a clear and fast way to evaluate whether a multiplex network can potentially exhibit super-diffusion. See Figs.~\ref{fig:ERER}, \ref{fig:SFSF} and \ref{fig:ERSF} from the Supplementary Material for the corresponding equivalent of Fig.~\ref{fig:test} for the duplex network configurations in Table~\ref{table:test}.

Overall, our predictions with parameter~$\eta$ perform well, except in the regions in which~$\eta$ is close to zero. The difference in the predictions can be explained in terms of the minimum strength as follows. First, note that the strength of any node of the ASN is the semi-sum of the strengths (degrees if the layers are unweighted) of the same node in the layers, $s_i^{\mathcal{A}}=(s_i^{[1]}+s_i^{[2]})/2$. In PC configurations, the node with minimum strength of the ASN is the same as the node with minimum strength in both layers (e.g., node~1 in Fig.~\ref{fig:diagram}), thus its strength is $s_\text{min}^{\mathcal{A}}=(s_\text{min}^{[1]} + s_\text{min}^{[2]})/2$. Since this is an intermediate value between the minimum strengths of the layers, then $\eta<0$, except when $s_\text{min}^{[1]}=s_\text{min}^{[2]}$, for which $\eta=0$. This means that, in a PC configuration, $\eta$ cannot be positive, thus restricting the appearance of super-diffusion to the cases in which the minimum strengths of the layers are equal, or very close (due to the fluctuations in the second smallest eigenvalue). For ER-ER duplex networks this happens when $\langle k^{[1]} \rangle \approx \langle k^{[2]} \rangle$, i.e., close to the diagonal, as seen in Fig.~\ref{fig:test}.

The other extreme case corresponds to NC configurations, for which the minimum strength in one layer is coupled with the maximum strength in the other, e.g., the first and the last nodes in Fig.~\ref{fig:diagram}. If $s^{[1]}_{\min}>s^{[2]}_{\max}$ or $s^{[2]}_{\min}>s^{[1]}_{\max}$, we have that $\eta<0$, and the multiplex network can only exhibit super-diffusion if the difference between the minimum strengths of the layers is small enough. This condition correctly identifies the region where the super-diffusion is restricted; however, its confining precision depends greatly on the strength distribution, see Fig.~\ref{fig:delta} from the Supplementary Material.

Although this method looks quite naive, it performs pretty well in the scenarios often used as a benchmark to explain the super-diffusion phenomenon \cite{cencetti2019diffusive,cai2023occurrence}, especially when the strength distribution across the nodes has high variation. It not only has the advantage of being independent of the layer structures but also provides a justification for why NC configurations increment and PC configurations decrease the dynamic performance of multiplex networks observed in previous literature \cite{kumar2017mirror,cai2023occurrence}.

\section{Analytical bounds}
\label{sec:bounds}

The previous approach works best when the strength distribution of the nodes across the network is heterogeneous, but we also want to know what happens for more homogeneous strength distributions, e.g., when the layers of the multiplex network are random regular (RR) networks, for which all nodes have the same strength.

Ideally, solving the super-diffusion problem for any multiplex network would involve expressing $\lambda_2$ as a function of some structural parameters, $N$, $\langle k \rangle$, $s_{\min}$, or others, and the corresponding strength distribution for each layer and the multiplex network. Then, we could compute analytically the structural conditions for super-diffusion.

While this sounds unrealistic for any given multiplex network structure, some results exist for $\lambda_2$ in some specific networks. In particular, for large ($N \gg 1$) unweighted RR networks of degree~$k$ \cite{mckay1981expected},
\begin{equation} \label{RR}
  \lambda^\text{RR}_2(k)
  \approx
  k - 2\sqrt{k-1}\,.
\end{equation}
And for large unweighted ER networks \cite{jamakovic2008robustness},
\begin{equation} \label{ER}
  \begin{split}
  \lambda^\text{ER}_2(N,p)
  \approx
  p (N - 1) &
  - \sqrt{2p(1 - p)(N - 1) \log N}
  \\
  & + \sqrt{\frac{(N - 1)p(1 - p)}{2 \log N}} \log \sqrt{2 \pi \log \left( \frac{N^2}{2\pi} \right)}
  \\
  & - \sqrt{\frac{(N - 1)p(1 - p)}{2 \log N}}\,\gamma\,,
\end{split}
\end{equation}
where $p$ is the connection probability of the nodes and $\gamma$ is the Euler–Mascheroni constant.

With these expressions, we have the RR and ER layers covered; however, the expression for the multiplex network is still missing. Suppose a duplex network formed by two unweighted RR layers with degree $k^{[\alpha]}$ and RA inter-layer connections. The probability of having a link overlap in the ASN, i.e., having a weight $w_{ij}^{\mathcal{A}}=1$, is given by
\begin{equation}
    p(w_{ij}^{\mathcal{A}} = 1) = p(w_{ij}^{[1]} = 1)\, p(w_{ij}^{[2]} = 1) = \frac{(k^{[1]})^2}{2M^{[1]} - 1}\, \frac{(k^{[2]})^2}{2M^{[2]} - 1}\,,
\end{equation}
where $k^{[\alpha]}$ are the node degrees of each layer, and $M^{[\alpha]}$ the number of links of each layer. Since $M^{[\alpha]} = N k^{[\alpha]} / 2$,
\begin{equation}
    p(w_{ij}^{\mathcal{A}} = 1) \approx \frac{k^{[1]} k^{[2]}}{N^2}\,.
\end{equation}
For sparse networks this probability of overlap goes to zero. The ASN Laplacian
\begin{equation}
    L^{\mathcal{A}} = S^{\mathcal{A}} - \frac{1}{2} (W_1+W_2) = \frac{1}{2} \overline{L}\,,
\end{equation}
where $\overline{L} = 2S^{\mathcal{A}} - (W_1+W_2)$, can be interpreted as the Laplacian matrix of an unweighted RR network with parameter $k=(k^{[1]}+k^{[2]}) = 2s^{\mathcal{A}}$, being $s^{\mathcal{A}}$ the strength of each node in the ASN. This stems from the fact that, when there is no overlap between the layers, the ASN includes all the links in both layers, but with half the weight. Therefore, for the duplex network we have that
\begin{equation}
    \lambda^{\mathcal{A}}_2(s^{\mathcal{A}}) \approx  \frac{1}{2} \lambda^\text{RR}_2(2s^{\mathcal{A}}),
\end{equation}
which can be generalized to the multiplex case by
\begin{equation} \label{RR-m}
    \lambda^{\mathcal{A}}_2(s^{\mathcal{A}},M) \approx  \frac{1}{M} \lambda^\text{RR}_2(Ms^{\mathcal{A}}) .
\end{equation}
This argument is also valid for the case of a multiplex network formed by unweighted ER layers with link probability $p^{[\alpha]}$ and RA inter-layer connections:
\begin{equation} \label{ER-m}
    \lambda^{\mathcal{A}}_2(N,p^{\mathcal{A}},M) \approx  \frac{1}{M} \lambda^\text{ER}_2(N,Mp^{\mathcal{A}}) .
\end{equation}
where $p^{\mathcal{A}} = \frac{1}{M}\sum_{\alpha} p^{[\alpha]}$.

With this, we can now bound the super-diffusion region by finding the values of the parameters that, according to Eqs.~\eqref{RR} and~\eqref{RR-m} for RR layers, or Eqs.~\eqref{ER} and~\eqref{ER-m} for ER layers, make $\xi = 0$.

\begin{figure}[tb!]
\begin{tabular}[t]{ll}
  \multicolumn{1}{c}{\footnotesize ER-ER (RA)} & \multicolumn{1}{c}{\footnotesize RR-RR (RA)}
  \\
  \includegraphics[width=0.48\textwidth]{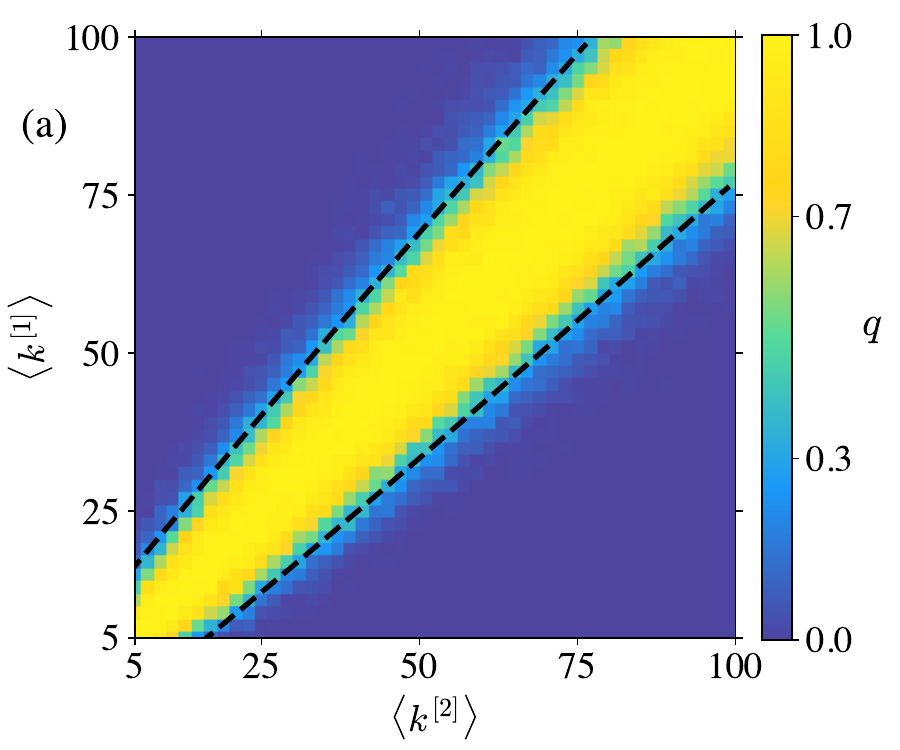}\phantomsubcaption\label{fig:bound:a} &      \includegraphics[width=0.48\textwidth]{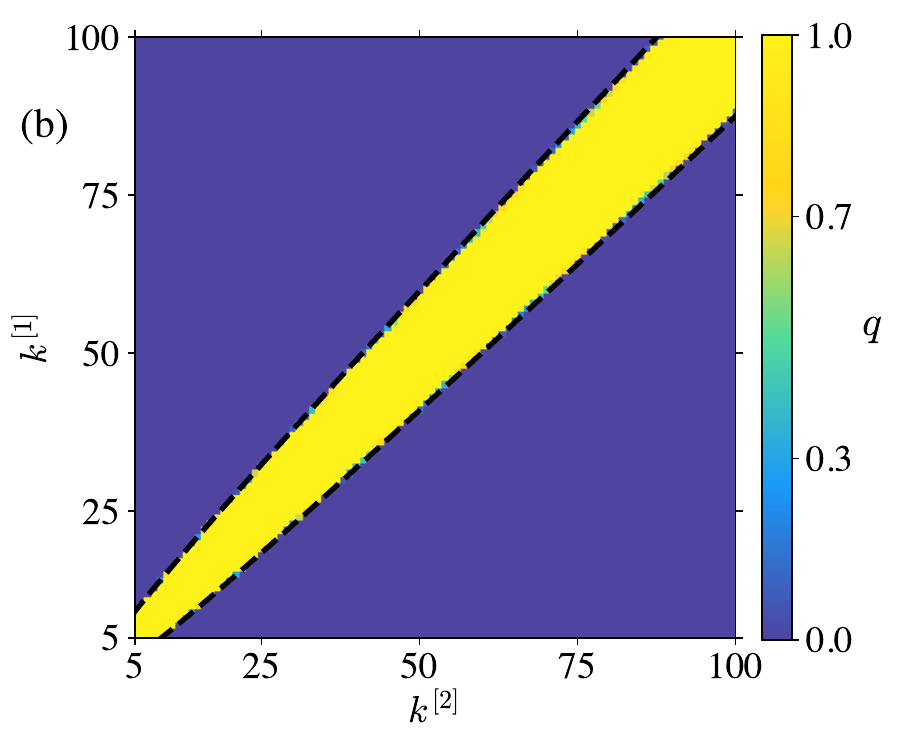}\phantomsubcaption\label{fig:bound:b}
  \\
  \includegraphics[width=0.445\textwidth]{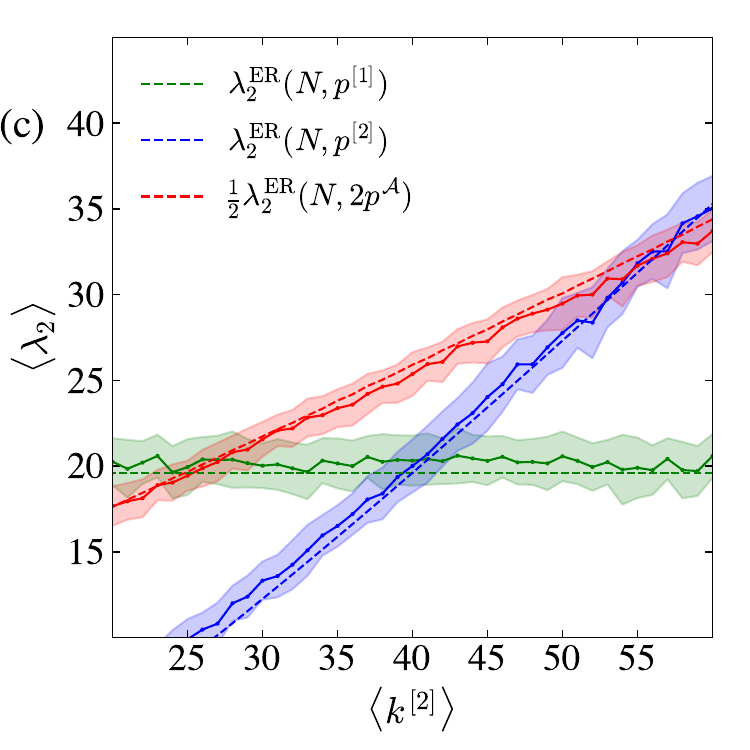}\phantomsubcaption\label{fig:bound:c} &      \includegraphics[width=0.445\textwidth]{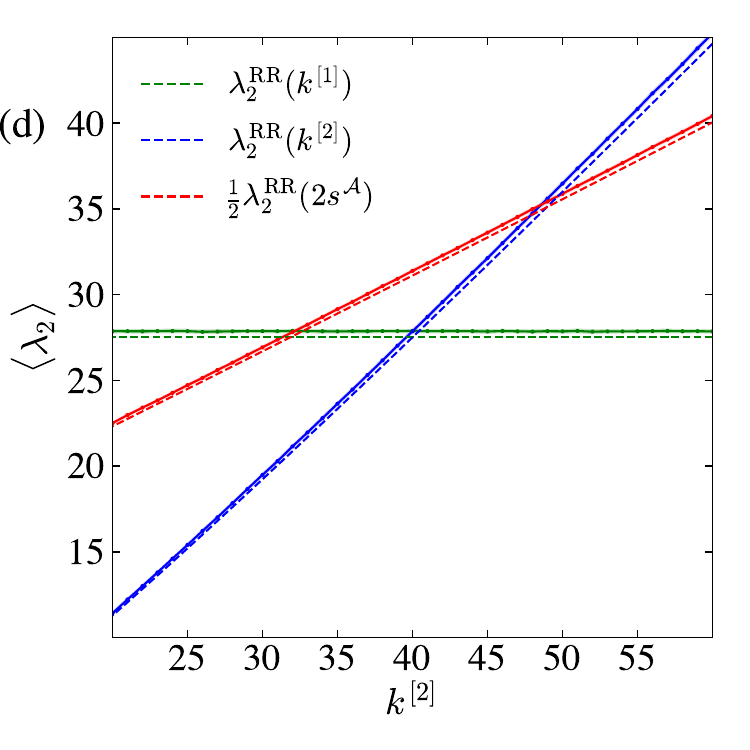}\phantomsubcaption\label{fig:bound:d}
\end{tabular}
\caption{Super-diffusion probability of RA duplex networks with (a) ER-ER and (b) RR-RR layers, all with $N=1000$. Black dashed lines represent the proposed theoretical bounds. Average second smallest eigenvalue (and bands spanning one standard deviation) for (c) ER-ER (RA) networks with $p^{[1]}=40/(N-1)$, and (d) RR-RR (RA) networks with $k^{[1]}=40$. Solid lines correspond to numerical results, and dashed lines to the theoretical curves. Color green used for layer~1, blue for layer~2, and red for the duplex network. We compute 50~different duplex networks for each combination of the structural parameters and ensure that the layers are connected networks.}
\end{figure}

In Figs.~\ref{fig:bound:a} and~\ref{fig:bound:b}, we represent the super-diffusion probability of duplex networks for the cases of two unweighted ER layers and two unweighted RR layers, respectively, together with the proposed theoretical bounds for each case. These bounds are found by solving $\xi = 0$, with the corresponding theoretical approximations of the second smallest eigenvalues, using the Newton-Raphson method. We observe a perfect agreement between the numerical results and the proposed bounds, validating the approach and providing for the first time an analytical justification for the super-diffusion occurrence.

Moreover, we can see in Fig.~\ref{fig:bound:c} that, since the variability of $\lambda_2$ for ER networks is relatively wide, there is a gradual transition between the super-diffusive and the non-super-diffusive regimes around the theoretical bounds, unlike for RR networks, for which the transitions are sharp and very close to the theoretical bounds, see Fig.~\ref{fig:bound:d}. This phenomenon further justifies the results previously shown in Fig.~\ref{fig:test}.

As we have seen, for bounding the super-diffusion region, it is not only important the structure of the layers but also the resulting ASN. Thus, we can bound the duplex network formed by unweighted ER layers with NC inter-layer connections because, when interconnecting the nodes in this fashion, we compensate the upper deviations from the mean value $\langle k^{[1]} \rangle$ with the lower deviations of $ \langle k^{[2]} \rangle$, and vice versa. As a result, in the duplex network, all the nodes will end up with a value of $s_i^{\mathcal{A}}$ close to $(\langle k^{[1]} \rangle + \langle k^{[2]} \rangle)/2$, i.e., a RR network with this strength value.

\begin{figure}[tb!]
\begin{tabular}[t]{ll}
     \includegraphics[width=0.49\textwidth]{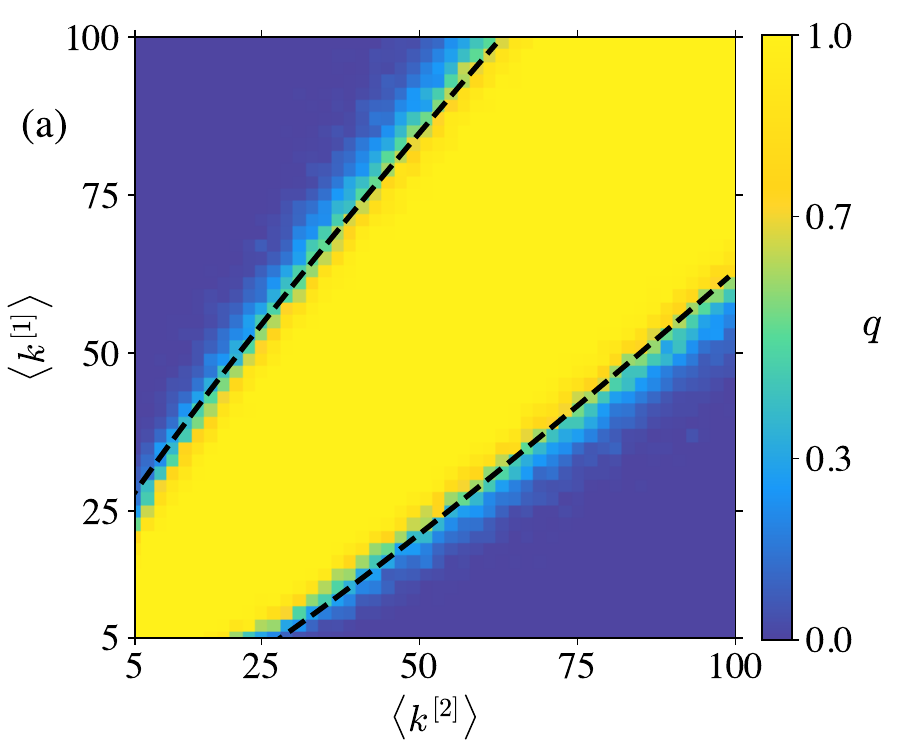}\phantomsubcaption\label{fig:nc:a} &      \includegraphics[width=0.41\textwidth]{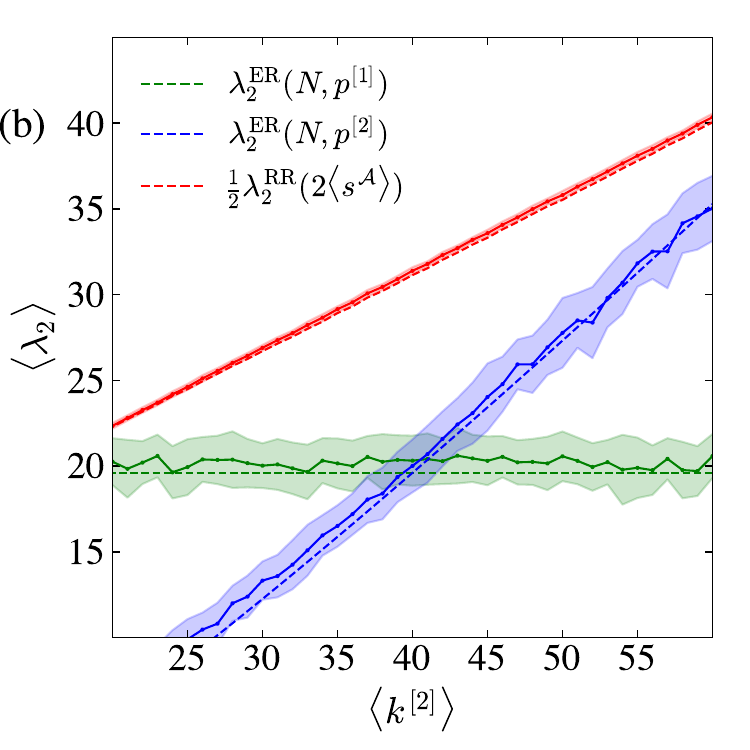}\phantomsubcaption\label{fig:nc:b}
\end{tabular}
\caption{(a) Super-diffusion probability of NC duplex networks with ER-ER layers, all with $N=1000$. Black dashed lines represent the proposed theoretical bounds based on approximating the ASN with a RR network. (b) Average second smallest eigenvalue (and bands spanning one standard deviation) for ER-ER (NC) networks with $p^{[1]}=40/(N-1)$. Solid lines correspond to numerical results, and dashed lines to the theoretical curves. Color green used for layer~1, blue for layer~2, and red for the duplex network. We compute 50~different duplex networks for each combination of the structural parameters and ensure that the layers are connected networks.}
\label{fig:nc}
\end{figure}

In Fig.~\ref{fig:nc} we verify the effectiveness of this argument by solving $\xi=0$ using the $\lambda_2$ of the corresponding RR superposition instead of the ER approximation. We see an almost perfect bounding of the super-diffusion region in Fig.~\ref{fig:nc:a}, and that the numerical results align with the corresponding theoretical curves in Fig.~\ref{fig:nc:b}.

\section{Maximizing multiplex diffusion}
\label{sec:maximum}

At this point, all the results indicate that the minimum node strength of the network plays an important role in determining the value of $\lambda_2$. This relationship allows us to justify the results in previous literature, which have identified the NC configuration as an effective strategy for increasing the occurrence of super-diffusion \cite{cai2023occurrence}. With this high correlation between the node strengths, the inter-layer connection arrangement, and the dynamics, we hypothesize that the NC configuration is actually the best strategy in multiplex networks to induce super-diffusion.

To test this hypothesis, we have explored the space of possible inter-layer connections in a similar way as in \cite{serrano2017optimizing}. Our aim is to maximize the value of $\lambda_2^{\mathcal{A}}$ by changing the inter-layer connections, or equivalently, by permuting the nodes of one layer and keeping the nodes of the other layer fixed. For that, we use simulated annealing (SA) \cite{kirkpatrick1983optimization}, with an acceptance probability $\min(1,\exp(\Delta \lambda_2^{\mathcal{A}}/T_f))$, where $\Delta \lambda_2^{\mathcal{A}}$ is the variation of the second smallest eigenvalue of the ASN after an inter-layer connection change, and $T_f$ is the temperature, which is linearly annealed. The two key aspects of the optimization problem are the starting point and the proposed changes at each step. In general, a NC configuration has different possible arrangements when multiple nodes share the same strength, a possibility previously overlooked. Since we observed that a random NC configuration is already a good starting point, we will consider two procedures: \textit{restricted~SA}, in which we maximize $\lambda_2^{\mathcal{A}}$ starting with a random NC configuration and proposing changes of inter-layer connections only with nodes of the same strength, and \textit{free~SA}, in which we start with a random NC configuration and all changes of inter-layer connections are allowed.

\begin{figure}[tb!]
\centering
\includegraphics[scale=.6]{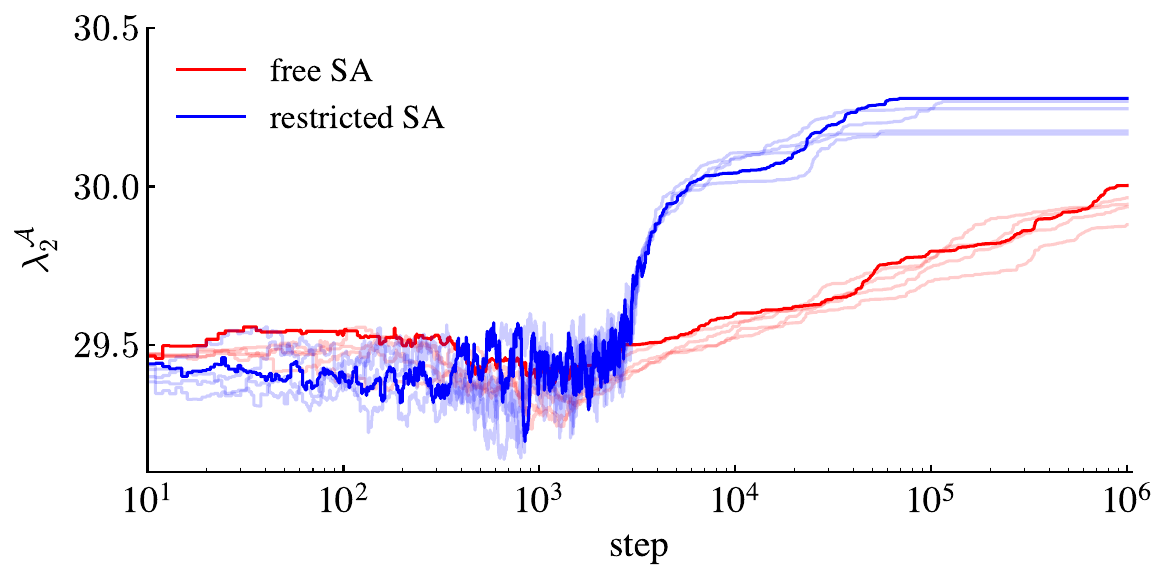}
\caption{Second smallest eigenvalue as a function of the SA steps for a duplex network formed by two ER layers with $N=500$, $k^{[1]} = 25$, and $k^{[2]} = 50$, chosen to be close to the super-diffusion boundary.}
\label{fig:SA}
\end{figure}

\begin{figure}[tb!]
\centering
\includegraphics[scale=.6]{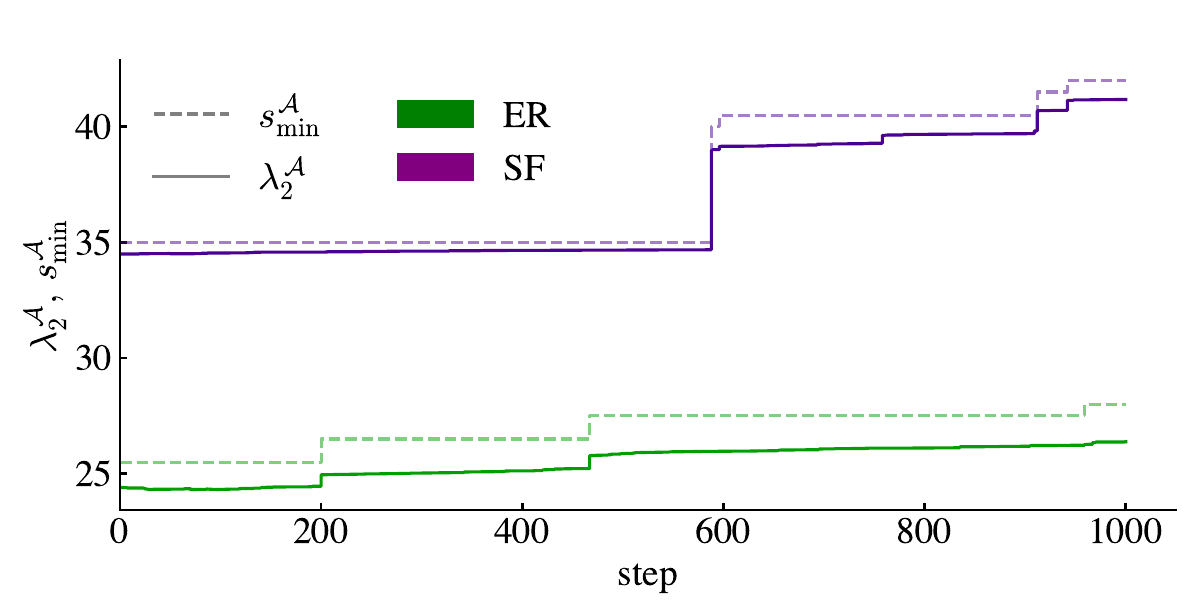}
\caption{Evolution of the second smallest eigenvalue (continuous lines) and minimum node strength (dashed lines) of duplex networks in a free~SA optimization, starting from RA inter-layer connections. One duplex is formed by two ER layers (green lines), and the other is formed by SF layers (purple lines), both with $N = 500$, $k^{[1]} = 25$, and $k^{[2]} = 50$.}
\label{fig:SMIN}
\end{figure}

In Fig.~\ref{fig:SA}, we observe various realizations of the optimization process for a duplex network formed by two ER layers, highlighting the best one for each case. We can see that restricted~SA is consistently better than free~SA, indicating that preserving the NC configuration is important to maximize duplex diffusion. Moreover, we can observe that a random NC configuration without optimization always outperforms the results from \cite{serrano2017optimizing} and from any free~SA optimization starting from a RA inter-layer connection, suggesting that maximizing $s^{\mathcal{A}}_{\min}$ is far more important than maximizing layer dissimilarity. This is supported by the observations that in a free~SA, starting from a RA inter-layer connection, the major increments of $\lambda^{\mathcal{A}}_2$ are always from changes that increment the value of $s^{\mathcal{A}}_{\min}$, as shown in Fig.~\ref{fig:SMIN}.

Finally, we have performed a more exhaustive optimization by considering all the duplex networks in Fig.~\ref{fig:nc:a} with restricted~SA, to see whether we could make the super-diffusion region wider than a NC configuration with random connections over nodes with the same strength. We have only been able to induce super-diffusion in less than $1\%$ of the duplex networks, all situated in the region where $\xi$ is close to zero (i.e., those close to the boundaries), supporting our hypothesis that a NC configuration is the best way to induce super-diffusion, and that minor improvements can be made by optimization within the NC configurations. Additionally, we repeated the analysis for the duplex networks in Fig.~\ref{fig:bound:b}, and the same was obtained; only the duplex networks that fall within the boundaries have a slight potential of inducing super-diffusion when rearranging inter-layer connections. See Figs.~\ref{fig:ER-opt} and \ref{fig:RR-opt} from the Supplementary Material to visualize the region where we could induce super-diffusion in both cases.

\begin{figure}[tb!]
\centering
\includegraphics[scale=1]{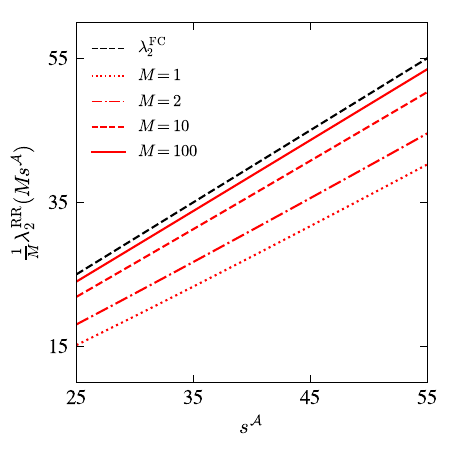}
\caption{Theoretical second-smallest eigenvalue of multiplex networks formed by unweighted RR layers as a function of $s^{\mathcal{A}}$ for different values of $M$. The black dashed line corresponds to Eq.~\eqref{fc}.}
\label{fig:limitRR}
\end{figure}

All the above results have further implications for the modification and design of network systems for better dynamical properties. The most immediate one is that, to enhance diffusion, it is better to have strength regularity across the network, not only in the sense of a more homogeneous strength distribution across the nodes but also within a node with fixed strength; having more links with less weight is better than having fewer links with a higher weight. This last observation is supported by Fig.~\ref{fig:limitRR}, which shows that Eq.~\eqref{RR-m} monotonically increases with the number of layers $M$, and as the ASN tends to a fully connected network, the $\lambda_2^{\mathcal{A}}$ approaches the upper bound outlined in Eq.~\eqref{fidler}.

\section{Conclusions}
\label{sec:conclusions}

In this work, we extend our current knowledge of the interplay between the structure and dynamics of complex networks by investigating the super-diffusion phenomenon in multiplex networks. First, we introduce a structural parameter $\eta$ to predict if a given multiplex structure can potentially induce super-diffusion. We then test this parameter on typical network structures, obtaining good prediction accuracy for duplex networks composed of ER and SF layers. Moreover, this parameter can justify the previous literature where NC inter-layer connections were observed to be more prone to induce super-diffusion and, in general, better dynamical properties in multiplex networks.

We also introduce a novel framework to bound the region of super-diffusion for some specific multiplex structures. We validate this approach by testing it on duplex networks composed of unweighted ER networks with both RA and NC inter-layer connections, as well as on unweighted RR networks.

Finally, we propose that a NC inter-layer connection is the most effective strategy for promoting super-diffusion in multiplex networks. This is supported in both the formulation of the structural parameter and the optimizations conducted with simulated annealing. We observe that preserving a NC configuration in the changes of inter-layer connections consistently leads to lower diffusion times. On top of that, we note a pattern in the free optimization process: significant increments in $\lambda_2^{\mathcal{A}}$ are consistently followed by increases in $s^{\mathcal{A}}_{\min}$, thereby supporting the formulation of $\eta$ and strengthening the proposal.

In summary, we have introduced novel insights into the super-diffusion phenomenon of multiplex networks and how it is highly related to node strength. These findings have potential implications for general network dynamics and open new questions, such as how other properties can reduce diffusion times when different nodes share the same strength.

\section*{Declaration of competing interest}

The authors declare that they have no known competing financial interests or personal relationships that could have appeared to influence the work reported in this paper.

\section*{Data availability}

No data was used for the research described in the article.

\section*{Acknowledgments}

L.T.\ acknowledges financial support from Universitat Rovira i Virgili (2023PMF-PIPF-21).
This work was supported by Ministerio de Ciencia e Innovaci\'on (PID2021-128005NB-C21, RED2022-134890-T), MICIU/AEI/10.13039/501100011033 FEDER EU (PID2022-142600NB-I00), Generalitat de Catalunya (2021SGR-633), and Universitat Rovira i Virgili (2023PFR-URV-00633).


\renewcommand*{\thefigure}{{S\arabic{figure}}}
\renewcommand{\figurename}{{Fig.}}
\setcounter{figure}{0}

\clearpage

\begin{center}
    {\Large Structural prediction of super-diffusion in multiplex networks} \\~\\
    \large{\textit{Supplementary Material}} \\~\\

    {\sc Llu\'is Torres-Hugas, Jordi Duch, Sergio G\'omez} \\~\\

    Departament d'Enginyeria Inform\`{a}tica i Matem\`{a}tiques \\
    Universitat Rovira i Virgili \\ 43007 Tarragona, Spain
\end{center}

\clearpage

\begin{figure}[tb!]
\includegraphics[scale=0.65]{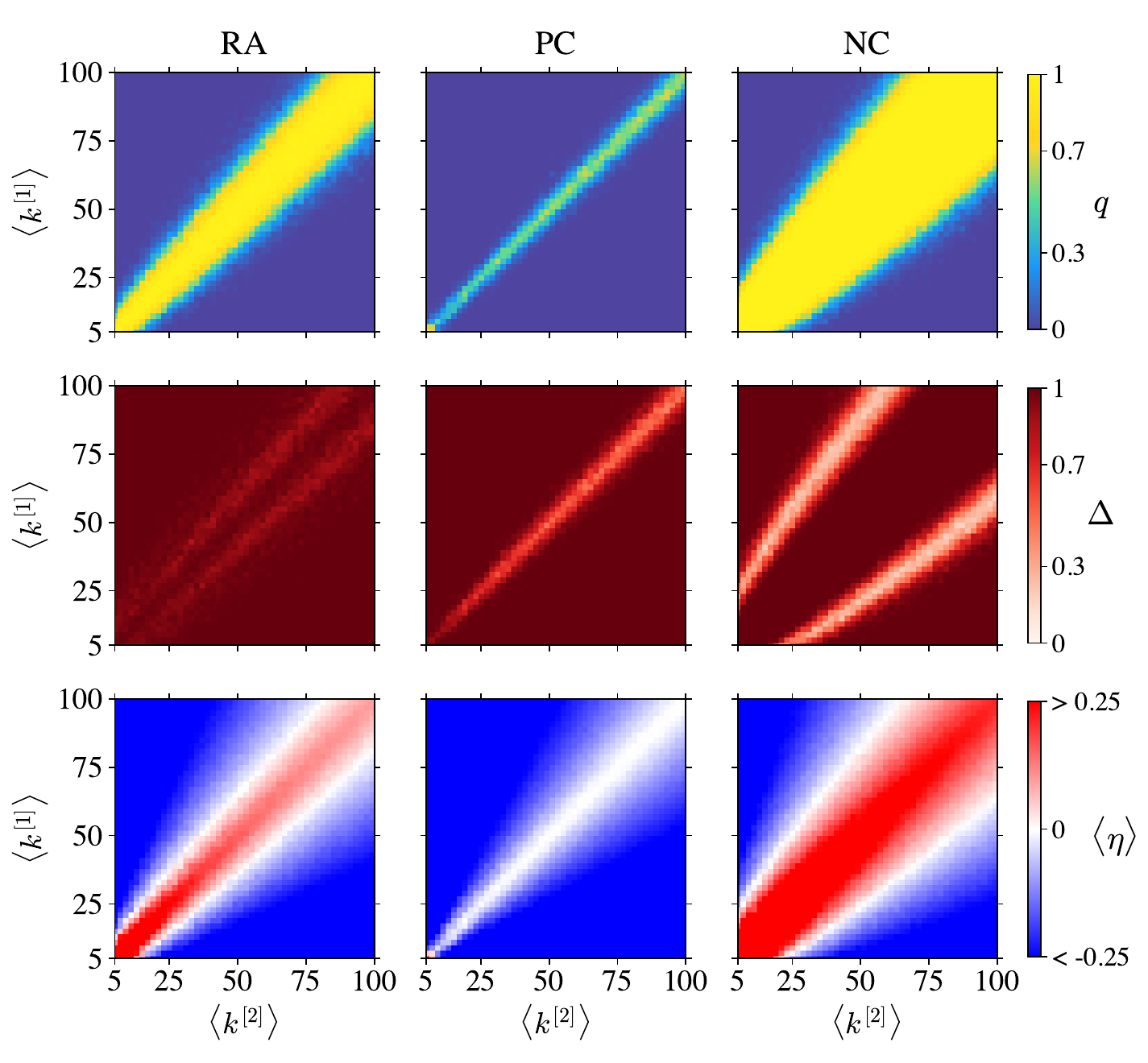}
\caption{Super-diffusion probability~$q$ (top row), prediction accuracy~$\Delta$ (middle row), and average structural parameter~$\langle \eta \rangle$ (bottom row) as a function of the average degrees of the layers, for duplex networks composed by two unweighted ER layers and with $N=500$. Each column corresponds to one of the three different models of inter-layer connection: RA, PC, and NC. We compute 50~different duplex networks for each combination of $p^{[1]}$ and $p^{[2]}$, with these connection probabilities ranging from~$5/(N-1)$ to~$100/(N-1)$ in steps of~$1/(N-1)$, and ensuring that the layers are connected networks.}
\label{fig:ERER}
\end{figure}

\clearpage

\begin{figure}[tb!]
\includegraphics[scale=0.65]{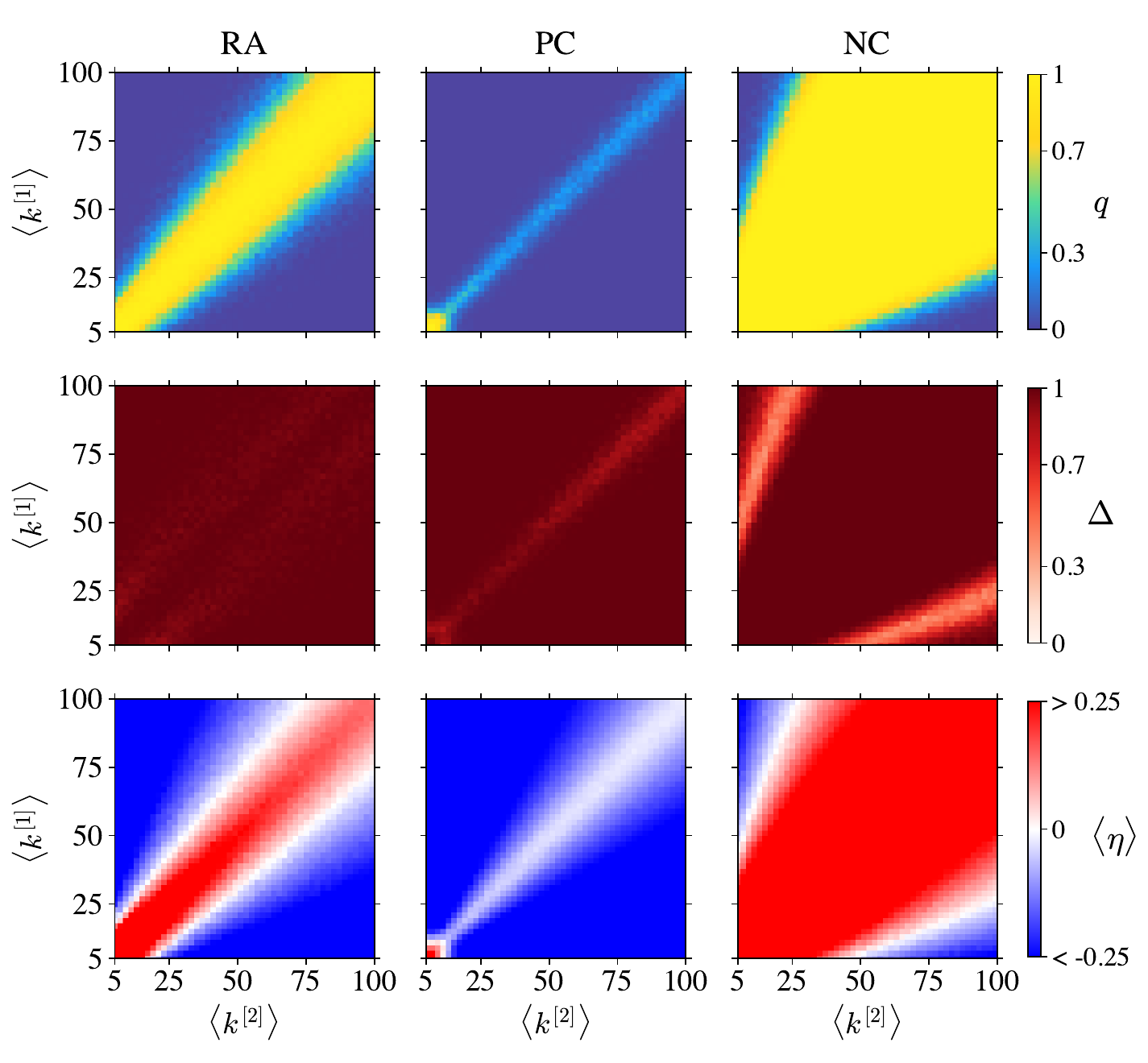}
\caption{Super-diffusion probability~$q$ (top row), prediction accuracy~$\Delta$ (middle row), and average structural parameter~$\langle \eta \rangle$ (bottom row) as a function of the average degrees of the layers, for duplex networks composed by two unweighted SF layers and with $N=500$. Each column corresponds to one of the three different models of inter-layer connection: RA, PC, and NC. We compute 50~different duplex networks for each combination of $M^{[1]}$ and $M^{[2]}$, with these number of links ranging from~$5N$ to~$100N$ in steps of~$N$, and ensuring that the layers are connected networks.}
\label{fig:SFSF}
\end{figure}

\clearpage

\begin{figure}[tb!]
\includegraphics[scale=0.65]{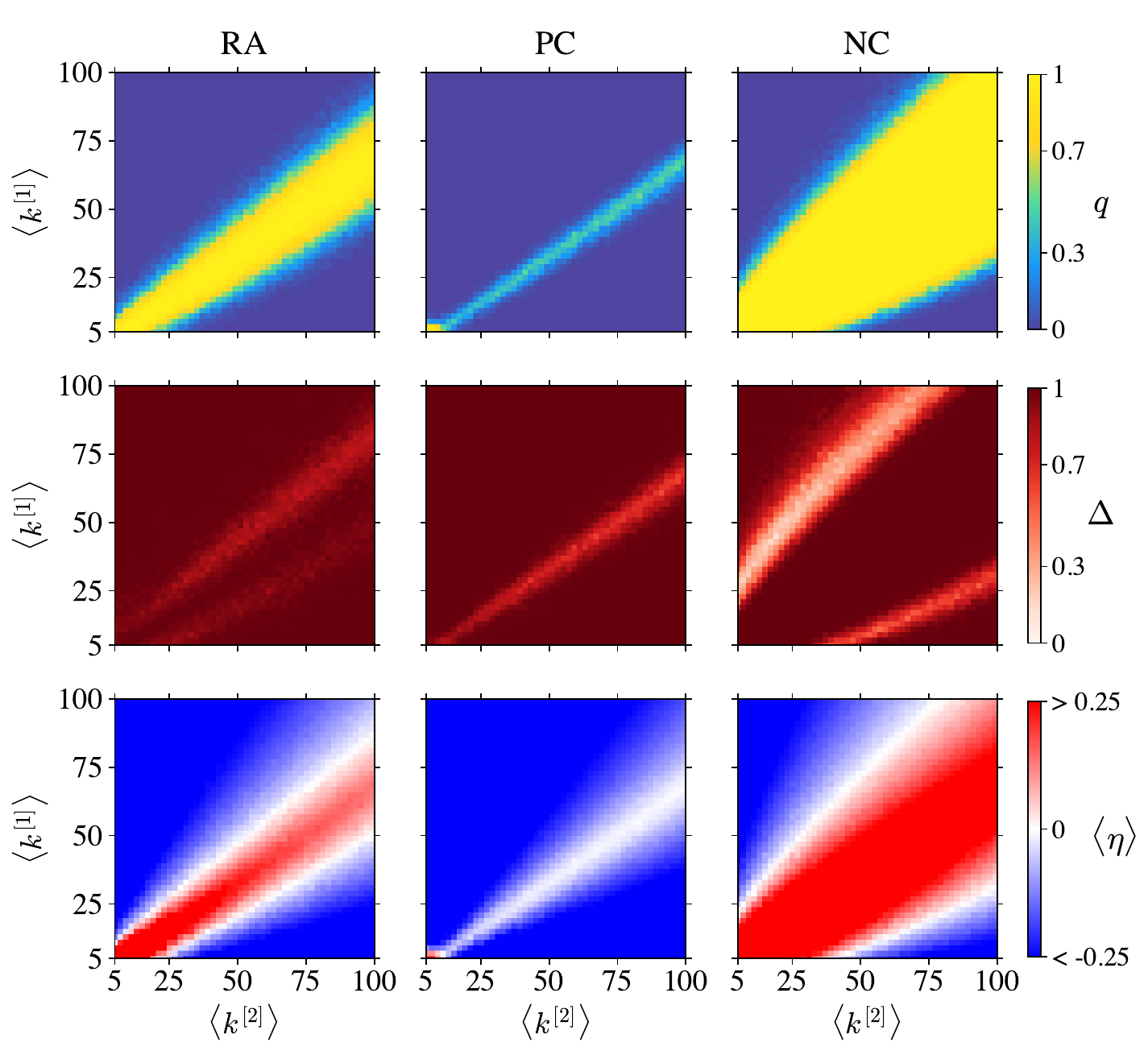}
\caption{Super-diffusion probability~$q$ (top row), prediction accuracy~$\Delta$ (middle row), and average structural parameter~$\langle \eta \rangle$ (bottom row) as a function of the average degrees of the layers, for duplex networks composed by one unweighted ER layer and one unweighted SF layer with $N=500$. Each column corresponds to one of the three different models of inter-layer connection: RA, PC, and NC. We compute 50 different duplex networks for each combination of connection probability~$p^{[1]}$ ranging from~$5/(N-1)$ to~$100/(N-1)$ in steps of~$1/(N-1)$ and number of links~$M^{[2]}$ ranging from~$5N$ to~$100N$ in steps of~$N$, and ensuring that the layers are connected networks.}
\label{fig:ERSF}
\end{figure}

\clearpage

\begin{figure}[tb!]
\includegraphics[scale=0.68]{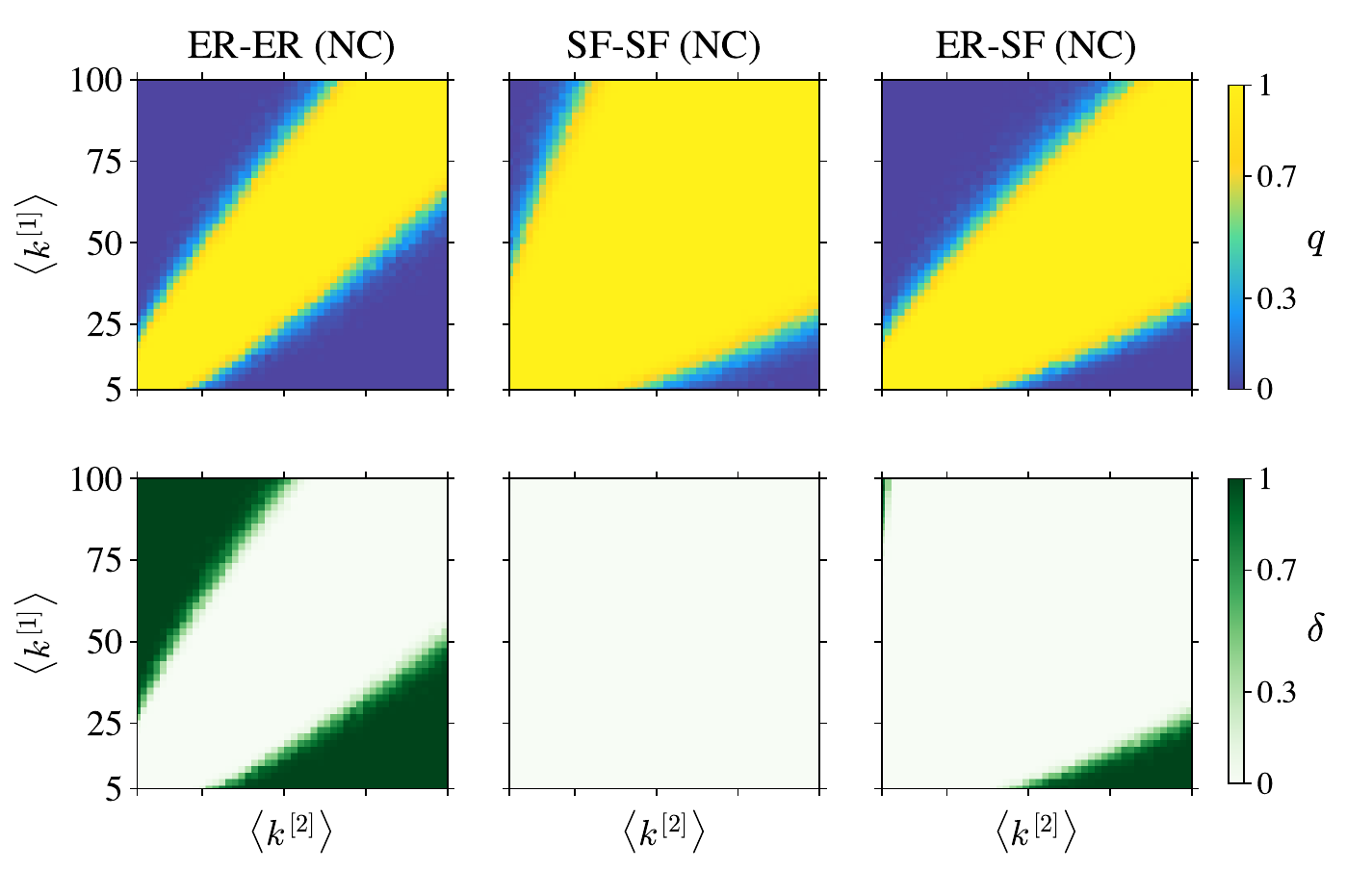}
\caption{Super-diffusion probability~$q$ (top row), and ratio of duplex networks that verify the restricting super-diffusion condition~$\delta$: $s^{[1]}_{\min}>s^{[2]}_{\max}$ or $s^{[2]}_{\min}>s^{[1]}_{\max}$ (bottom row) as a function of the average degrees of the layers, for duplex networks with $N=500$. Each column corresponds to one of the three different configurations: ER-ER (NC), SF-SF (NC), and ER-SF (NC). We compute 50~different duplex networks for each combination of structural parameters and ensure that the layers are connected networks. Note that, for SF networks, this condition is hard to accomplish because of the broad degree distribution.}
\label{fig:delta}
\end{figure}

\clearpage

\begin{figure}
    \centering
    \includegraphics[scale=1]{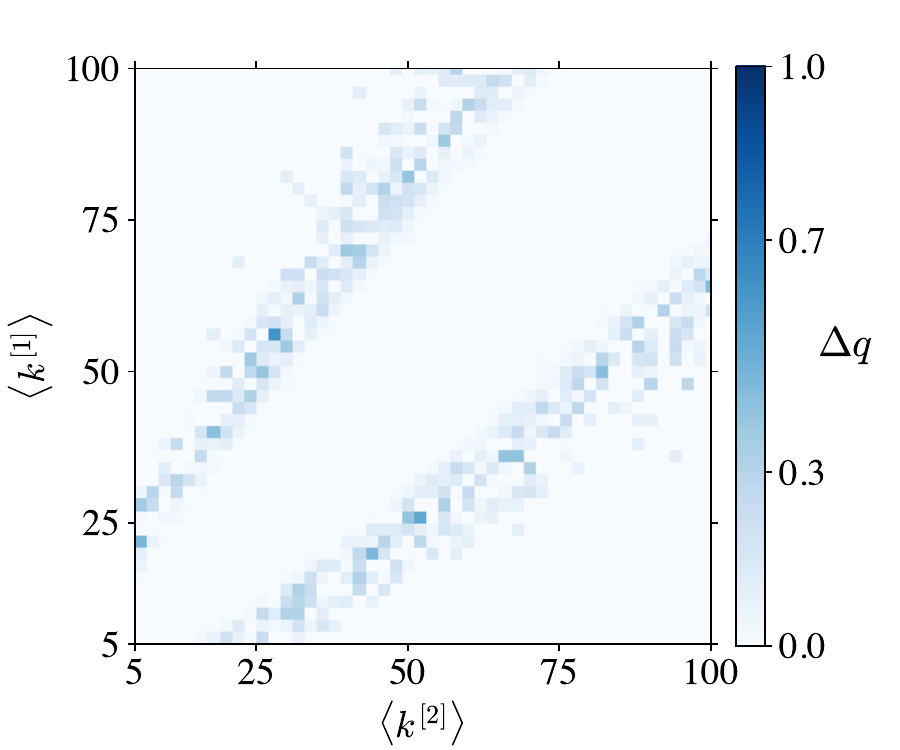}
    \caption{Super-diffusion probability increment after a restricted SA optimization process of $10^3$ steps on each duplex networks as a function of their average connectivity. Duplex networks formed by two unweighted ER layers with $N=1000$. We compute 50 different duplex networks for each combination of structural parameters and ensure that the layers are connected networks. Inter-layer connections are NC.}
    \label{fig:ER-opt}
\end{figure}

\clearpage

\begin{figure}
    \centering
    \includegraphics[scale=1]{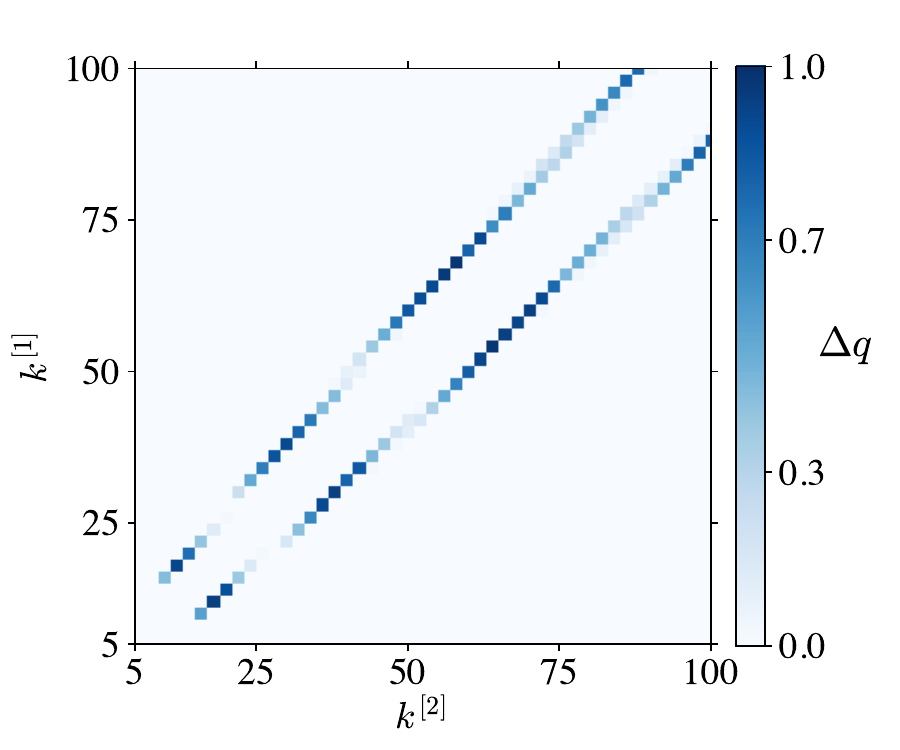}
    \caption{Super-diffusion probability increment after a restricted SA optimization process of $10^3$ steps on each duplex networks as a function of their average connectivity. Duplex networks formed by two unweighted RR layers with $N=1000$. We compute 50 different duplex networks for each combination of structural parameters and ensure that the layers are connected networks.}
    \label{fig:RR-opt}
\end{figure}

\end{document}